\documentclass[aps,prl,twocolumn,groupedaddress,showpacs,showkeys]{revtex4}
\usepackage{graphicx}
\begin{document}
\title{Cavity QED with degenerate atomic levels and
polarization-degenerate field mode}

\author{V.A.
Reshetov}\email[E-mail:~]{vareshetov@tltsu.ru}\affiliation{Togliatti
State University, Belorusskaya str., 14, Togliatti, 445020, Russia}

\date{\today}

\begin{abstract}
The Jaynes-Cummings model with degenerate atomic levels and
polarization-degenerate field mode is considered. The general
expression for the system evolution operator is derived. The
analytical expressions for such operators in the case of low values
($J \leq 3/2$) of atomic angular momentum are obtained. The
polarization properties of the photon emitted into the cavity by an
excited atom are studied with an account of relaxation processes for
arbitrary angular momenta of atomic levels.
\end{abstract}

\pacs{42.50.Dv} \keywords{Jaynes-Cummings model, polarization
properties, degenerate levels}

\maketitle

\section{Introduction}

The original Jaynes-Cummings model \cite{n1} and its numerous
descendants \cite{n2} constitute basic theoretical tools in quantum
optics (see, for example, the textbooks \cite{n3}-\cite{n5} and the
reviews  \cite{n6}-\cite{n9}). In the present article one more
extension of Jaynes-Cummings model, taking into account the
degeneracy of atomic levels and polarization degeneracy of the field
mode, is considered. The reason for such extension is that the
levels of isolated atoms in the micro-cavity are degenerate in the
projections of the angular momentum on the quantization axis (unless
this degeneracy is somehow lifted), while the single frequency mode
of the quantized field in the cavity may obtain two independent
polarizations in the plane perpendicular to the cavity axis (unless
the symmetry in the cavity is somehow broken). For example, in the
experiments \cite{n10}-\cite{n14} the field mode was on resonance
with the transition between Rydberg states of $^{85}Rb$ atoms with
the angular momenta of atomic levels $J=3/2,5/2$, while in the
experiments \cite{n15}-\cite{n22} the field mode was on resonance
with the transitions between the hyperfine-structure components with
the angular momenta $F=1,2,3$ of electronic levels $5S_{1/2}$ and
$5P_{3/2}$ of $^{85}Rb$ and $^{87}Rb$ atoms. On the other hand the
degeneracy of atomic levels may be employed to control the
interaction of the atom with the field mode by means of atomic
preparation in some special states, while the control of the field
polarization in the cavity may provide tools for handling the
photon-polarization qubits. The Jaynes-Cummings model with
degenerate atomic levels but with the fixed polarization of the
field mode was considered in \cite{n23}-\cite{n25} with an
application to the theory of one-atom maser. However for the
applications of atom-cavity systems as universal nodes of quantum
networks \cite{n20} the two polarization states of the photon in the
cavity should be taken into account. The various aspects of the
Jaynes-Cummings model with degenerate atomic levels and
polarization-degenerate field mode were considered previously in
\cite{n26}-\cite{n30}. In the present article the general solution
of such a model, based on the conservation laws of energy and
projection of the angular momentum on the quantization axis, is
proposed. In the first section the basics of the model are described
and the decomposition of the space of system states into the
subspaces invariant under the action the system hamiltonian is
performed. It is shown that the problem of diagonalization of the
hamiltonian is reduced to the diagonalization of hermitian matrix of
the dimension equal to the degree of degeneracy of atomic levels. In
the second section the evolution operator of the system is
considered and an example of the dynamics of an atom with the
resonant transition $J_{0}=3\rightarrow J_{1}=4$ in the unpolarized
thermal field is studied numerically. In the third section the
analytical expressions for the eigenvalues and eigenvectors of the
system hamiltonian are presented for the low values ($J_{0},J_{1}
\leq 3/2$) of atomic angular momentum. In the fourth section the
polarization dynamics of the photon stored in the cavity is
discussed. In the fifth section the polarization properties of the
photon emitted into the cavity by an excited atom is studied with an
account of relaxation processes.

\section{The model basics}

The model system consists of a single two-level atom with degenerate
levels and a single polarization-degenerate mode of the quantized
electromagnetic field, the basis states of the system space being
the product of atomic and field states $$ |J_{a},m_{a}\rangle
|n_{1},n_{2}\rangle.$$ The atomic states
     \begin{equation}\label{q1}
|J_{a},m_{a}\rangle
     \end{equation}
are characterized by the number $a = 0,1$ of the atomic energy level
$E_{a}$, $a=0$ refers to the ground state, $a=1$ refers to the
excited state. The frequency of the atomic transition: $\omega_{0} =
(E_{1} - E_{0})/\hbar $, $E_{1} = \hbar\omega_{0}/2$, $E_{0} = -
\hbar\omega_{0}/2$, $J_{a}$ is the total angular momentum of atomic
level $E_{a}$, $m_{a}$ is the projection of this angular momentum on
the quantization axis Z. The quantization axis Z is directed along
the micro-cavity axis. The states of the field mode
     \begin{equation}\label{q2}
|n_{1},n_{2}\rangle .
     \end{equation}
are characterized by the number of photons $n_{k}$ with the
polarization $\textbf{s}_{k}$, $k = 1,2$, where $\textbf{s}_{k}$ are
the two orthonormal vectors in the XY plane:
$\textbf{s}_{i}\textbf{s}_{j}^{*} = \delta_{ij}$, $i,j = 1,2$,
$n=n_{1}+n_{2}$ being the total photon number.

The hamiltonian of the model system
     \begin{equation}\label{q3}
\hat{H} = \hat{H_{0}} + \hat{V}
     \end{equation}
is the sum of the free system hamiltonian
     \begin{equation}\label{q4}
\hat{H_{0}} = \hbar\omega \hat{n} + \frac{1}{2} \hbar\omega_{0}
\left(\hat{P}_{A}^{(1)} - \hat{P}_{A}^{(0)}\right),
     \end{equation}
and the interaction operator
     \begin{equation}\label{q5}
\hat{V} = - \left(\hat{\textbf{d}} + \hat{\textbf{d}}^{\dag}\right)
\hat{\textbf{E}}.
     \end{equation}
Here
     \begin{equation}\label{q6}
\hat{P}_{A}^{(a)} = \sum_{m_{a} = - J_{a}}^{J_{a}}
|J_{a},m_{a}\rangle \langle J_{a},m_{a}|,
     \end{equation}
is the projection operator on the subspace of atomic states
referring to the level $a$,
     \begin{equation}\label{q7}
\hat{\textbf{d}} = \sum_{m_{0},m_{1}} (\textbf{d})^{01}_{m_{0}m_{1}}
|J_{0},m_{0}\rangle \langle J_{1},m_{1}|,
     \end{equation}
is the operator of the electric dipole moment transition $J_{1}
\rightarrow J_{0}$, $(\textbf{d})^{01}_{m_{0}m_{1}}$ being the
matrix elements of this operator, $\omega$ is the frequency of the
field mode, the electric field operator
     \begin{equation}\label{q8}
\hat{\textbf{E}} = \hat{\textbf{e}} + \hat{\textbf{e}}^{\dag},~~~
\hat{\textbf{e}} = i e_{0} (\textbf{s}_{1} \hat{a}_{1} +
\textbf{s}_{2} \hat{a}_{2} ),
     \end{equation}
is expressed through the annihilation $\hat{a}_{k}$ and creation
$\hat{a}_{k}^{\dag}$ operators of the photon with the polarization
$\textbf{s}_{k}$, $\hat{n}_{k} = \hat{a}_{k}^{\dag}\hat{a}_{k}$
being the number operator of such photons, $\hat{n} = \hat{n}_{1} +
\hat{n}_{2}$,
     \begin{equation}\label{q9}
e_{0} = \sqrt{\frac{\hbar \omega}{2\varepsilon_{0} V}},
     \end{equation}
$V$ is the mode volume, the atom is placed at the center of the
cavity $z=0$.

In the rotating wave approximation the hamiltonian (\ref{q3}) may be
written as follows:
     \begin{equation}\label{q10}
\hat{H} = \hbar\omega \hat{h}_{0} - \frac{\hbar}{2} \hat{\Omega},~~~
\hat{h}_{0} = \hat{n} + \frac{1}{2} \left(\hat{P}_{A}^{(1)} -
\hat{P}_{A}^{(0)}\right),
     \end{equation}
     \begin{equation}\label{q11}
\hat{\Omega} = \Delta \left(\hat{P}_{A}^{(0)} -
\hat{P}_{A}^{(1)}\right) + \theta \left(\hat{G} +
\hat{G}^{\dag}\right),
     \end{equation}
     \begin{equation}\label{q12}
\hat{G} = -i \left(\hat{G}_{1} + \hat{G}_{2}\right),~~~ \hat{G}_{k}
= \hat{g}_{k}\hat{a}_{k}^{\dag}~~~ \hat{g}_{k} =
\hat{\textbf{g}}\textbf{s}_{k}^{*}.
     \end{equation}
     \begin{equation}\label{q13}
\Delta = \omega_{0} - \omega,~~~ \theta = 2de_{0}/\hbar,~~~
\hat{\textbf{g}} = \hat{\textbf{d}}/d,
     \end{equation}
$d$ being the reduced matrix element of the electric dipole moment
operator of the transition $J_{1} \rightarrow J_{0}$.

Let us choose the two orthonormal polarization vectors
$\textbf{s}_{1}$ and $\textbf{s}_{2}$ to stand for $\sigma^{+}$ and
$\sigma^{-}$ polarizations correspondingly, the circular components
of these vectors being
     \begin{equation}\label{q14}
s_{1q} = \delta_{q,-1},~~~s_{2q} = \delta_{q,+1}.
     \end{equation}
With such a choice the field states (\ref{q2}) may be defined by the
total photon number $n$ and by the projection $\sigma$ of the photon
angular momentum on the quantization axis $Z$:
     \begin{equation}\label{q15}
|n,\sigma \rangle ,
     \end{equation}
where $n = n_{1} + n_{2}$, $\sigma = n_{1} - n_{2}$, $n_{1}$ is the
number of the photons with the $\sigma^{+}$ polarization
$\textbf{s}_{1}$, $n_{2}$ is the number of the photons with the
$\sigma^{-}$ polarization $\textbf{s}_{2}$. At the given photon
number $n$ the projection $\sigma$ attains the following $(n+1)$
values:
     \begin{equation}\label{q16}
\sigma = - n + 2k,~~~ k = 0,1,2,...,n.
     \end{equation}
Then the action of operators $\hat{G}_{1}$ and $\hat{G}_{2}$
(\ref{q12}) on the system states is as follows:
     \begin{equation}\label{q17}
\hat{G}_{1}^{\dag} |J_{0},m \rangle |n,\sigma \rangle  =
G_{1,mn\sigma}|J_{1},m+1 \rangle |n-1,\sigma -1 \rangle,
     \end{equation}
     \begin{equation}\label{q18}
\hat{G}_{2}^{\dag} |J_{0},m \rangle |n,\sigma \rangle  =
G_{2,mn\sigma} |J_{1},m-1 \rangle |n-1,\sigma +1 \rangle.
     \end{equation}
The matrix elements $G_{1,mn\sigma}$ and $G_{2,mn\sigma}$ of these
operators are expressed through Wigner $3j$-symbols:
     \begin{equation}\label{q19}
G_{1,mn\sigma}  = f_{1}(m)\sqrt{\frac{1}{2}(n+\sigma)},
     \end{equation}
     \begin{equation}\label{q20}
G_{2,mn\sigma}  = f_{2}(m)\sqrt{\frac{1}{2}(n-\sigma)},
     \end{equation}
     \begin{equation}\label{q21}
f_{1}(m) = (-1)^{J_{0}-m} \left(\matrix{J_{0}&1&J_{1} \cr
-m&-1&m+1}\right),
     \end{equation}
     \begin{equation}\label{q22}
f_{2}(m) = (-1)^{J_{0}-m} \left(\matrix{J_{0}&1&J_{1} \cr
-m&1&m-1}\right).
     \end{equation}
As it follows from (\ref{q17})-(\ref{q18}) the interaction operator
$\hat{\Omega}$ (\ref{q11}) does not change the excitation number
$n=p+a$, where $p$ is the photon number and $a=0,1$ is the number of
the atomic level, and it does not change the projection $l=m+\sigma$
of the system total angular momentum on the quantization axis $Z$.
Consequently the space $V$ of the system states may be decomposed
into a set of subspaces $V^{(nl)}$ with given numbers $n$ and $l$
which are invariant under the action of the interaction operator
$\hat{\Omega}$:
     \begin{equation}\label{q23}
V = \bigcup_{n=0}^{\infty}\bigcup_{l=-L}^{L} V^{(nl)},~~~
\hat{\Omega} V^{(nl)} = V^{(nl)},
     \end{equation}
where $L=J_{0}+n$ is the maximum value of the system total angular
momentum projection on the quantization axis at given excitation
number $n$. The interaction operator $\hat{\Omega}$ may be expanded
into the sum of operators acting invariantly in subspaces
$V^{(nl)}$:
     \begin{equation}\label{q24}
\hat{\Omega} = \sum_{n=0}^{\infty}\sum_{l=-L}^{L}
\hat{\Omega}^{(nl)},~~~ \hat{\Omega}^{(nl)} =\hat{P}^{(nl)}
\hat{\Omega} \hat{P}^{(nl)},
     \end{equation}
where $\hat{P}^{(nl)}$ is the projection operator on the subspace
$V^{(nl)}$. The subspaces $V^{(nl)}$ may be further decomposed into
two subspaces:
     \begin{equation}\label{q25}
V^{(nl)} = V^{(0,n,l)}\bigcup V^{(1,n-1,l)},
     \end{equation}
where $V^{(anl)}$ is the subspace of states with the atom at level
$a=0,1$, photon number $n$ and $m_{a}+\sigma =l$. The orthonormal
basis of the subspace $V^{(anl)}$ consists of the states:
     \begin{equation}\label{q26}
|w^{(anl)}_{k}\rangle = |J_{a},m_{k}\rangle
|n,l-m_{k}\rangle,~m_{k}=m^{(anl)}_{min}+2k,
     \end{equation}
$k = 0,1,...,N^{(anl)}-1$, where $m^{(anl)}_{min}$ is the minimum
value of $m_{a}$ at given $n$ and $l$, $N^{(anl)}$ is the dimension
of the subspace $V^{(anl)}$. The subspace $V^{(anl)}$ is invariant
under the action of operator $\hat{D}_{a}$, where $\hat{D}_{0} =
\hat{G}\hat{G}^{\dag}$, $\hat{D}_{1} = \hat{G}^{\dag}\hat{G}$, while
$\hat{G}$ is defined by (\ref{q12}). Instead of the orthonormal
basis (\ref{q26}) in the subspace $V^{(anl)}$ let us consider the
orthonormal basis consisting of the eigenvectors of the hermitian
operator $\hat{D}_{a}$ acting in this subspace. Next, the subspace
$V^{(anl)}$ may be decomposed into two subspaces:
     \begin{equation}\label{q27}
V^{(anl)} = V^{(anl)}_{d}\bigcup V^{(anl)}_{c},
     \end{equation}
where the orthonormal basis $|d^{(anl)}_{k}\rangle$,
$k=0,1,...,N^{(anl)}_{d}-1$, of the the subspace $V^{(anl)}_{d}$
consists of the eigenvectors of the operator $\hat{D}_{a}$ with zero
eigenvalues, while the orthonormal basis $|v^{(anl)}_{k}\rangle$,
$k=0,1,...,N^{(anl)}_{c}-1$
($N^{(anl)}_{d}+N^{(anl)}_{c}=N^{(anl)}$), of the subspace
$V^{(anl)}_{c}$ consists of the eigenvectors of the operator
$\hat{D}_{a}$ with real positive non-zero eigenvalues
$[\xi^{(anl)}_{k}]^{2}$:
     \begin{equation}\label{q28}
\hat{D}_{a}|v^{(anl)}_{k}\rangle = [\xi^{(anl)}_{k}]^{2}
|v^{(anl)}_{k}\rangle.
     \end{equation}
If the states $|v^{(0,n,l)}_{k}\rangle$ in the subspace
$V^{(0,n,l)}_{c}$ are the orthonormal eigenvectors of the operator
$\hat{D}_{0}$ with non-zero eigenvalues:
     \begin{equation}\label{q29}
[\xi^{(0,n,l)}_{k}]^{2}=[\xi^{(nl)}_{k}]^{2},
     \end{equation}
then the states
     \begin{equation}\label{q30}
|v^{(1,n-1,l)}_{k}\rangle = \frac{1}{\xi^{(nl)}_{k}}
\hat{G}^{\dag}|v^{(0,n,l)}_{k}\rangle
     \end{equation}
are the orthonormal eigenvectors of the operator $\hat{D}_{1}$ in
the subspace $V^{(1,n-1,l)}_{c}$ with non-zero eigenvalues:
     \begin{equation}\label{q31}
[\xi^{(1,n-1,l)}_{k}]^{2} = [\xi^{(0,n,l)}_{k}]^{2} =
[\xi^{(nl)}_{k}]^{2}.
     \end{equation}
So, the subspace
     \begin{equation}\label{q32}
V^{(nl)}_{c} = V^{(0,n,l)}_{c}\bigcup V^{(1,n-1,l)}_{c}
     \end{equation}
may be decomposed into $N^{(nl)}_{c} = N^{(0,n,l)}_{c} =
N^{(1,n-1,l)}_{c}$ two-dimensional subspaces
     \begin{equation}\label{q33}
V^{(nl)}_{ck} = V^{(0,n,l)}_{ck}\bigcup V^{(1,n-1,l)}_{ck},
     \end{equation}
($k=0,1,...,N^{(nl)}_{c}-1$) which are invariant under the action of
interaction operator $\hat{\Omega}$. The orthonormal basis in the
subspace $V^{(nl)}_{ck}$ consists of two states
$|v^{(0,n,l)}_{k}\rangle $ and $|v^{(1,n-1,l)}_{k}\rangle$. In this
subspace:
     \begin{equation}\label{q34}
\hat{G}^{\dag} |v^{(0,n,l)}_{k}\rangle = \xi^{(nl)}_{k}
|v^{(1,n-1,l)}_{k}\rangle,
     \end{equation}
     \begin{equation}\label{q35}
\hat{G} |v^{(1,n-1,l)}_{k}\rangle = \xi^{(nl)}_{k}
|v^{(0,n,l)}_{k}\rangle .
     \end{equation}
The eigenvalues and eigenvectors of the interaction operator
$\hat{\Omega}$ in the subspace $V^{(nl)}_{ck}$ are as follows:
     \begin{equation}\label{q36}
\hat{\Omega} |u^{(nl,\pm)}_{k}\rangle = \pm \Omega^{(nl)}_{k}
|u^{(nl,\pm)}_{k}\rangle ,
     \end{equation}
     \begin{equation}\label{q37}
|u^{(nl,\pm)}_{k}\rangle = c^{(nl,\pm)}_{k} |v^{(0,n,l)}_{k}\rangle
\pm c^{(nl,\mp)}_{k} |v^{(1,n-1,l)}_{k}\rangle,
     \end{equation}
     \begin{equation}\label{q38}
\Omega^{(nl)}_{k} = \sqrt{\Delta^{2} + \theta^{2}\xi^{(nl)2}_{k}},
     \end{equation}
     \begin{equation}\label{q39}
c^{(nl,\pm)}_{k} =\sqrt{\frac{1}{2}\left(1 \pm
\frac{\Delta}{\Omega^{(nl)}_{k}}\right)}.
     \end{equation}
Consequently the interaction operator $\hat{\Omega}^{(nl)}$
(\ref{q24}) in the subspace $V^{(nl)}$ may be presented in such a
way:
     \begin{equation}\label{q40}
\hat{\Omega}^{(nl)} = \hat{\Omega}^{(nl)}_{d} +
\hat{\Omega}^{(nl)}_{c},
     \end{equation}
     \begin{equation}\label{q41}
\hat{\Omega}^{(nl)}_{d} = \Delta \left(\hat{P}^{(0,n,l)}_{d} -
\hat{P}^{(1,n-1,l)}_{d} \right),
     \end{equation}
     \begin{equation}\label{q42}
\hat{\Omega}^{(nl)}_{c}  = \sum_{k=0}^{N^{(nl)}_{c}-1}
\Omega^{(nl)}_{k} \left(\hat{P}^{(nl,+)}_{ck} -
\hat{P}^{(nl,-)}_{ck} \right),
     \end{equation}
where
     \begin{equation}\label{q43}
\hat{P}^{(anl)}_{d} = \sum_{k=0}^{N^{(anl)}_{d}-1}
|d^{(anl)}_{k}\rangle \langle d^{(anl)}_{k}|
     \end{equation}
and
     \begin{equation}\label{q44}
\hat{P}^{(nl,\pm)}_{ck} = |u^{(nl,\pm)}_{k}\rangle \langle
u^{(nl,\pm)}_{k}|
     \end{equation}
are the projection operators on the subspaces $V^{(anl)}_{d}$ and
$V^{(nl,\pm)}_{ck}$ correspondingly.

So the problem of diagonalization of the interaction operator
$\hat{\Omega}^{(nl)}$ is reduced to the problem of diagonalization
of the $N^{(anl)}\times N^{(anl)}$ hermitian matrix
     \begin{equation}\label{q45}
D^{(anl)}_{k'k} = \langle w^{(anl)}_{k'}|\hat{D}_{a}
|w^{(anl)}_{k}\rangle ,
     \end{equation}
of the operator $\hat{D}_{a}$ in the subspace $V^{(anl)}$. The
non-zero elements of the matrix $D^{(anl)}_{k'k}$ are expressed
through the functions $G_{k,mn\sigma}$ defined by the formulae
(\ref{q19})-(\ref{q22}).

The minimum $m^{(anl)}_{min}$ and maximum $m^{(anl)}_{max}$ values
of the projection $m_{a}$ of the atom angular momentum on the
quantization axis at given $n$ and $l$, and the dimension $N^{(anl)}
= 1+(m_{max} - m_{min})/2$ of the subspace $V^{(anl)}$ may be found
from the conditions:
     \begin{equation}\label{q46}
|m_{k}|\leq J_{a},~ |l-m_{k}|\leq n,~ m_{k} = m^{(anl)}_{min} + 2k,
     \end{equation}
where $k = 0,1,...,N^{(anl)} - 1$. Since the projection of the
photon angular momentum $\sigma$ varies with the step 2 (\ref{q16})
the two sets of subspaces $V^{(anl)}_{p}$ ($p=0,1$) may be
distinguished with
     \begin{equation}\label{q47}
l = - J_{a} - n + p + 2s,~s = 0,1,...,L^{(an)}_{p},
     \end{equation}
where
     \begin{equation}\label{q48}
L^{(an)}_{p} = J^{(a)}_{p} + n,
     \end{equation}
     \begin{equation}\label{q49}
J^{(a)}_{p} = \left\{ \matrix{J_{a} - p,~ \textrm{for integer}~
J_{a} \cr J_{a} - \frac{1}{2},~ \textrm{for half-integer}~
J_{a}}\right\}.
     \end{equation}
Then we obtain from (\ref{q46})-(\ref{q47}):
     \begin{equation}\label{q50}
m_{min}^{(anlp)} = \left\{ \matrix{-J_{a}+p,~ 0\leq s \leq n \cr
-J_{a}+p + 2(s-n) ,n < s \leq L^{(an)}_{p}}\right\},
     \end{equation}
     \begin{equation}\label{q51}
N_{p}^{(anl)} = \left\{ \matrix{s+1,~ 0\leq s \leq s_{min}^{(anp)}
\cr s_{min}^{(anp)}+1 ,~ s_{min}^{(anp)} < s \leq s_{max}^{(anp)}
\cr L^{(an)}_{p}+1-s,~ s_{max}^{(anp)} < s \leq
L^{(an)}_{p}}\right\},
     \end{equation}
where
     \begin{equation}\label{q52}
s_{min}^{(anp)} = \min\{ n, J^{(a)}_{p}\},~ s_{max}^{(anp)} = \max\{
n, J^{(a)}_{p}\}.
     \end{equation}
The maximum dimension $N^{(anp)}_{m}$ of the subspaces
$V^{(anl)}_{p}$ is defined by the formula:
     \begin{equation}\label{q53}
N^{(anp)}_{m} = s_{min}^{(anp)}+1 \leq J^{(a)}_{p} + 1.
     \end{equation}

\begin{figure}[h] \center
\includegraphics[width=9cm]{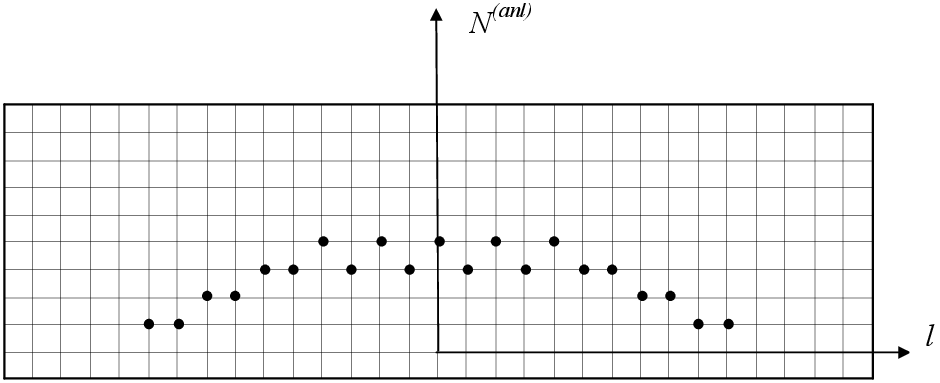}
\caption{The dependence of the subspace dimension $N^{(anl)}$ on $l$
for $J_{a}=3$, $n=7$.}\label{fig_1}
\end{figure}
\begin{figure}[h] \center
\includegraphics[width=9cm]{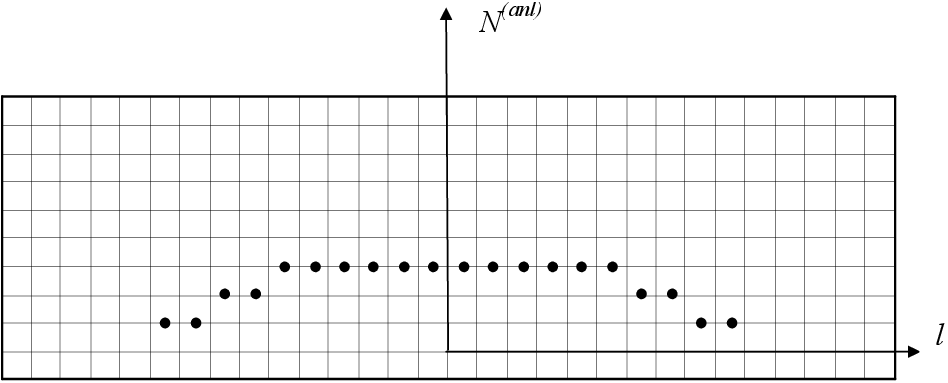}
\caption{The dependence of the subspace dimension $N^{(anl)}$ on $l$
for $J_{a}=5/2$, $n=7$.}\label{fig_2}
\end{figure}

The typical dependencies of the dimensions $N^{(anl)}$ of the
subspaces $V^{(anl)}$ on the projection $l$ of the system angular
momentum on the quantization axis $Z$ are presented in the Figures 1
and 2 in the cases of integer $J_{a}=3$ and half-integer $J_{a}=5/2$
values of atomic angular momenta correspondingly and the photon
number $n=7>J^{(a)}_{p}$.

\section{The system dynamics}

The system dynamics is described by the equation for its density
matrix $\hat{\sigma}$:
     \begin{equation}\label{q54}
\frac{d\hat{\rho}}{dt} = \frac{i}{2}
\left[\hat{\Omega},\hat{\rho}\right],~~~ \hat{\rho} = \hat{u}^{\dag}
\hat{\sigma} \hat{u},~~~ \hat{u} = e^{-i\omega t \hat{h}_{0}},
     \end{equation}
where $\hat{\rho}$ is the slowly-varying system density matrix. The
solution of the equation (\ref{q54}) is expressed through the
evolution operator $\hat{S}(t)$:
     \begin{equation}\label{q55}
\hat{\rho}(t) = \hat{S}(t)\hat{\rho}(0)\hat{S}^{\dag}(t),~~~
\hat{S}(t) = \exp \left\{\frac{i}{2} \hat{\Omega} t\right\}.
     \end{equation}
      \begin{equation}\label{q56}
\hat{S}(t) = \sum_{n=0}^{\infty}\sum_{l=-L}^{L} \hat{S}^{(nl)}(t),
     \end{equation}
     \begin{equation}\label{q57}
\hat{S}^{(nl)}(t) = \hat{P}^{(nl)}
\exp\left\{\frac{i}{2}\hat{\Omega}^{(nl)} t\right\}.
     \end{equation}
From the expressions (\ref{q40})-(\ref{q44}) for the interaction
operator we obtain the following expressions for the evolution
operator in terms of eigenvalues and eigenvectors of operators
$\hat{D}_{a}$ defined in the equations (\ref{q28})-(\ref{q39}):
     \begin{equation}\label{q58}
\hat{S}^{(nl)}(t) = \hat{S}^{(nl)}_{d}(t) + \sum_{a,b=0}^{1}
\hat{S}^{(nl)}_{ab}(t),
     \end{equation}
     \begin{equation}\label{q59}
\hat{S}^{(nl)}_{d}(t) = e^{i\Delta t/2} \hat{P}^{(0,n,l)}_{d} +
e^{-i\Delta t/2} \hat{P}^{(1,n-1,l)}_{d},
     \end{equation}
     \begin{equation}\label{q60}
\hat{S}^{(nl)}_{00}(t) = \sum_{k=0}^{N^{(nl)}_{c}-1} C^{(nl)}_{k}(t)
|v^{(0nl)}_{k}\rangle \langle v^{(0nl)}_{k}|,
     \end{equation}
     \begin{equation}\label{q61}
\hat{S}^{(nl)}_{11}(t) = \sum_{k=0}^{N^{(nl)}_{c}-1}
[C^{(nl)}_{k}(t)]^{*} |v^{(1,n-1,l)}_{k}\rangle \langle
v^{(1,n-1,l)}_{k}|,
     \end{equation}
     \begin{equation}\label{q62}
\hat{S}^{(nl)}_{01}(t) = \sum_{k=0}^{N^{(nl)}_{c}-1} S^{(nl)}_{k}(t)
|v^{(0nl)}_{k}\rangle \langle v^{(1,n-1,l)}_{k}|,
     \end{equation}
     \begin{equation}\label{q63}
\hat{S}^{(nl)}_{10}(t) = \sum_{k=0}^{N^{(nl)}_{c}-1} S^{(nl)}_{k}(t)
|v^{(1,n-1,l)}_{k}\rangle \langle v^{(0nl)}_{k}|,
     \end{equation}
     \begin{equation}\label{q64}
C^{(nl)}_{k}(t) = \cos\left(\frac{\Omega^{(nl)}_{k} t}{2}\right) + i
\frac{\Delta}{\Omega^{(nl)}_{k}} \sin\left(\frac{\Omega^{(nl)}_{k}
t}{2}\right),
     \end{equation}
     \begin{equation}\label{q65}
S^{(nl)}_{k}(t) = i\frac{\theta\xi^{(nl)}_{k}}{\Omega^{(nl)}_{k}}
\sin\left(\frac{\Omega^{(nl)}_{k} t}{2}\right).
     \end{equation}

As an example let us consider the dynamics of population of the
atomic excited level
     \begin{equation}\label{q66}
n_{1}(t) = tr\left\{\hat{P}_{A}^{(1)}\hat{\rho}(t)\right\}
     \end{equation}
in the case when the atom is initially in the equilibrium ground
state and the field is in the equilibrium thermal state, so that the
initial system density matrix is as follows:
     \begin{equation}\label{q67}
\hat{\rho}(0) = \frac{\hat{P}_{A}^{(0)}}{2J_{0}+1}
\sum_{n=0}^{\infty}\frac{p_{n}}{n+1} \sum_{\sigma = -n}^{n}
|n,\sigma\rangle \langle n,\sigma|,
     \end{equation}
     \begin{equation}\label{q68}
p_{n} = \frac{n_{c}^{n}}{(1+n_{c})^{n+1}},
     \end{equation}
where $n_{c}$ is an average photon number. In this case we obtain
from (\ref{q55})-(\ref{q68}):
     \begin{equation}\label{q69}
n_{1}(t) = \frac{1}{2J_{0}+1} \sum_{n=0}^{\infty}\frac{p_{n}}{n+1}
\sum_{l = -(J_{0}+n)}^{J_{0}+n} F_{nl}(t),
     \end{equation}
     \begin{equation}\label{q70}
F_{nl}(t) = \sum_{k=0}^{N^{(nl)}_{c}-1}
\frac{\theta^{2}\xi^{(nl)2}_{k}}{\Omega^{(nl)2}_{k}}
\sin^{2}\left(\frac{\Omega^{(nl)}_{k} t}{2}\right).
     \end{equation}
\begin{figure}[h] \center
\includegraphics[width=9cm]{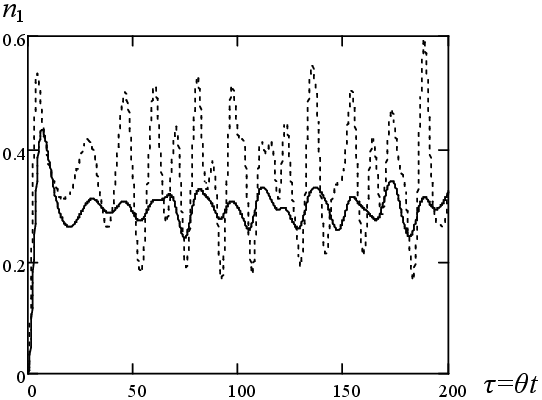}
\caption{The dependence of the population $n_{1}$ of the excited
atomic level on the dimensionless time $\tau = \theta t$ for the
transition $J_{0}=3\rightarrow J_{1}=4$, $\Delta =0.1 \theta$ and
$n_{c}=3$. The solid line refers to the thermally equilibrium
initial state of the system, while the dotted line refers to the
case, when the atom is initially in a "stretched" ground state
$|J_{0},J_{0}\rangle$ and only $\sigma^{+}$ mode of the thermal
field is present.}\label{fig_3}
\end{figure}
The dependence of the population $n_{1}$ of the excited atomic level
on the dimensionless time $\tau = \theta t$ is presented in the
Figure 3 for the transition $J_{0}=3\rightarrow J_{1}=4$, $\Delta
=0.1 \theta$ and $n_{c}=3$. The solid line refers to the case of the
equilibrium initial state of the system (\ref{q67}), while the
dotted line refers to the case of the system initial density matrix
     \begin{equation}\label{q71}
\hat{\rho}(0) = |J_{0},J_{0}\rangle \langle J_{0},J_{0}|
\sum_{n=0}^{\infty} p_{n} |n,n\rangle \langle n,n|,
     \end{equation}
when the atom is initially in a "stretched" ground state
$|J_{0},J_{0}\rangle$ and only $\sigma^{+}$ mode of the thermal
field (\ref{q68}) is present. In this case the original
Jaynes-Cummings model with two-level atom and single-mode field is
realised:
     \begin{equation}\label{q72}
n_{1}(t) = \sum_{n=0}^{\infty}p_{n}
\frac{\theta^{2}\xi_{n}^{2}}{\Omega^{2}_{n}}
\sin^{2}\left(\frac{\Omega_{n} t}{2}\right),
     \end{equation}
     \begin{equation}\label{q73}
\Omega_{n} = \sqrt{\Delta^{2} + \theta^{2}\xi^{2}_{n}},~ \xi_{n} =
\sqrt{\frac{n}{2J_{0}+3}} \delta_{J_{1},J_{0}+1}.
     \end{equation}

\section{The transitions with low values of atomic angular momentum}

The matrix elements of the evolution operator (\ref{q58}) are
expressed through the eigenvalues and eigenvectors of operators
$\hat{D}_{a}$ in the subspaces $V^{(anl)}$ (\ref{q59})-(\ref{q65}).
Since the dimensions of these subspaces do not exceed the value
$J_{a}+1$ (\ref{q53}) these eigenvalues and eigenvectors may be
easily obtained analytically in the case of low values of atomic
angular momentum $J_{a}\leq 3/2$. The explicit expressions of these
eigenvalues and eigenvectors for such transitions are presented in
this section.

\subsection{Transitions $J_{0}=0\rightarrow J_{1}=1$}

For the subspaces of the set $p=0$ (\ref{q47}) with
     \begin{equation}\label{q74}
l = - n + 2s,~s = 0,1,...,n,
     \end{equation}
at the values $s=0$ and $s=n$ ($l=\mp n$):
$N^{(0,n,l)}=N^{(1,n-1,l)}=N_{c}^{(nl)}=1$,
     \begin{equation}\label{q75}
|v^{(0,n,\pm n)}\rangle = |0,0\rangle |n,\pm n\rangle,~ \xi^{(nl)} =
\sqrt{n/3},
     \end{equation}
     \begin{equation}\label{q76}
|v^{(1,n-1,\pm n)}\rangle = i|1,\pm 1\rangle |n-1,\pm (n-1)\rangle ,
     \end{equation}
while at the values $s=1,...,n-1$: $N^{(0,n,l)}=N_{c}^{(nl)}=1$,
$N^{(1,n-1,l)}=2$, $N_{d}^{(1,n-1,l)}=1$,
     \begin{equation}\label{q77}
|v^{(0,n,l)}\rangle = |0,0\rangle |n,l\rangle,~ \xi^{(nl)} =
\sqrt{n/3},
     \end{equation}
     \begin{equation}\label{q78}
|v^{(1,n-1,\l)}\rangle = c_{0} |w_{0}^{(1,n-1,\l)}\rangle + c_{1}
|w_{1}^{(1,n-1,\l)}\rangle ,
     \end{equation}
     \begin{equation}\label{q79}
|d^{(1,n-1,\l)}\rangle = c_{1} |w_{0}^{(1,n-1,\l)}\rangle - c_{0}
|w_{1}^{(1,n-1,\l)}\rangle ,
     \end{equation}
where
     \begin{equation}\label{q80}
c_{0} = i \sqrt{\frac{n-l}{2n}},~ c_{1} = i \sqrt{\frac{n+l}{2n}},
     \end{equation}
     \begin{equation}\label{q81}
|w_{k}^{(1,n-1,l)}\rangle = |1,-1+2k \rangle |n-1,l+1-2k \rangle,~
k=0,1.
     \end{equation}

For the subspaces of the set $p=1$ (\ref{q47}) with
     \begin{equation}\label{q82}
l = - n + 1 + 2s,~s = 0,1,...,n-1,
     \end{equation}
$N^{(0,n,l)}=0$, $N^{(1,n-1,l)}=N_{d}^{(1,n-1,l)}=1$,
     \begin{equation}\label{q83}
|d^{(1,n-1,l)}\rangle = i|1,0\rangle |n - 1,l\rangle .
     \end{equation}

\subsection{Transitions $J_{0}=1\rightarrow J_{1}=0$}

For the subspaces of the set $p=0$ (\ref{q47}) with
     \begin{equation}\label{q84}
l = - (n+1) + 2s,~s = 0,1,...,n,
     \end{equation}
at the values $s=0$ and $s=n+1$ ($l=\mp (n+1)$):
$N^{(0,n,l)}=N_{d}^{(0,n,l)}=1$, $N^{(1,n-1,l)}=0$,
     \begin{equation}\label{q85}
|d^{(0,n,\pm (n+1))}\rangle = |1,\pm 1\rangle |n,\pm n\rangle,
     \end{equation}
while at the values $s=1,...,n$: $N^{(0,n,l)}=2$,
$N_{d}^{(0,n,l)}=1$, $N^{(1,n-1,l)}=N_{c}^{(nl)}=1$,
     \begin{equation}\label{q86}
|v^{(0,n,\l)}\rangle = c_{0} |w_{0}^{(0,n,\l)}\rangle + c_{1}
|w_{1}^{(0,n,\l)}\rangle ,
     \end{equation}
     \begin{equation}\label{q87}
|d^{(0,n,\l)}\rangle = c_{1} |w_{0}^{(0,n,\l)}\rangle - c_{0}
|w_{1}^{(0,n,\l)}\rangle ,
     \end{equation}
      \begin{equation}\label{q88}
|v^{(1,n-1,l)}\rangle = i|0,0\rangle |n-1,l\rangle,~ \xi^{(nl)} =
\sqrt{(n+1)/3},
     \end{equation}
where
     \begin{equation}\label{q89}
c_{0} =  \sqrt{\frac{n+1+l}{2(n+1)}},~ c_{1} =
\sqrt{\frac{n+1-l}{2(n+1)}},
     \end{equation}
     \begin{equation}\label{q90}
|w_{k}^{(0,n,l)}\rangle = |1,-1+2k \rangle |n,l+1-2k \rangle,~
k=0,1.
     \end{equation}
For the subspaces of the set $p=1$ (\ref{q47}) with
     \begin{equation}\label{q91}
l = - n + 2s,~s = 0,1,...,n,
     \end{equation}
$N^{(0,n,l)}=N_{d}^{(0,n,l)}=1$, $N^{(1,n-1,l)}=0$,
     \begin{equation}\label{q92}
|d^{(0,n,l)}\rangle = |1,0\rangle |n,l\rangle .
     \end{equation}

\subsection{Transitions $J_{0}=1\rightarrow J_{1}=1$}

For the subspaces of the set $p=0$ (\ref{q47}) with
     \begin{equation}\label{q93}
l = - (n+1) + 2s,~s = 0,1,...,n+1,
     \end{equation}
at the values $s=0$ and $s=n+1$ ($l=\mp (n+1)$):
$N^{(0,n,l)}=N_{d}^{(0,n,l)}=1$, $N^{(1,n-1,l)}=0$, the states
$|d^{(0,n,\pm (n+1))}\rangle$ are defined by the equation
(\ref{q85}), while at the values $s=1,...,n$: $N^{(0,n,l)}=2$,
$N_{d}^{(0,n,l)}=1$, $N^{(1,n-1,l)}=N_{c}^{(nl)}=1$,
     \begin{equation}\label{q94}
|v^{(0,n,\l)}\rangle = - c_{0} |w_{0}^{(0,n,\l)}\rangle + c_{1}
|w_{1}^{(0,n,\l)}\rangle ,
     \end{equation}
     \begin{equation}\label{q95}
|d^{(0,n,\l)}\rangle = c_{1} |w_{0}^{(0,n,\l)}\rangle + c_{0}
|w_{1}^{(0,n,\l)}\rangle ,
     \end{equation}
     \begin{equation}\label{q96}
|v^{(1,n-1,l)}\rangle = i|1,0\rangle |n-1,l\rangle,~ \xi^{(nl)} =
\sqrt{(n+1)/6},
     \end{equation}
where $c_{0}$, $c_{1}$ and the states $|w_{k}^{(0,n,l)}\rangle$ are
defined by the equations (\ref{q89})-(\ref{q90}).

For the subspaces of the set $p=1$ (\ref{q47}) with
     \begin{equation}\label{q97}
l = - n + 2s,~s = 0,1,...,n,
     \end{equation}
at the values $s=0$ and $s=n$ ($l=\mp n$):
$N^{(0,n,l)}=N^{(1,n-1,l)}=N_{c}^{(nl)}=1$,
     \begin{equation}\label{q98}
|v^{(0,n,\pm n)}\rangle = |1,0\rangle |n,\pm n\rangle,~ \xi^{(nl)} =
\sqrt{n/6},
     \end{equation}
     \begin{equation}\label{q99}
|v^{(1,n-1,\pm n)}\rangle = i|1,\pm 1\rangle |n-1,\pm (n-1)\rangle ,
     \end{equation}
while at the values $s=1,...,n-1$: $N^{(0,n,l)}=N_{c}^{(nl)}=1$,
$N^{(1,n-1,l)}=2$, $N_{d}^{(1,n-1,l)}=1$,
     \begin{equation}\label{q100}
|v^{(0,n,l)}\rangle = |1,0\rangle |n,l\rangle,~ \xi^{(nl)} =
\sqrt{n/6},
     \end{equation}
     \begin{equation}\label{q101}
|v^{(1,n-1,\l)}\rangle = c_{0} |w_{0}^{(1,n-1,\l)}\rangle - c_{1}
|w_{1}^{(1,n-1,\l)}\rangle ,
     \end{equation}
     \begin{equation}\label{q102}
|d^{(1,n-1,\l)}\rangle = c_{1} |w_{0}^{(1,n-1,\l)}\rangle + c_{0}
|w_{1}^{(1,n-1,\l)}\rangle ,
     \end{equation}
where $c_{0}$, $c_{1}$ and the states $|w_{k}^{(0,n,l)}\rangle$ are
defined by the equations (\ref{q80})-(\ref{q81}).

\subsection{Transitions $J_{0}=1/2\rightarrow J_{1}=1/2$}

For the subspaces of the set $p=0$ (\ref{q47}) with
     \begin{equation}\label{q103}
l = - (n+1/2) + 2s,~s = 0,1,...,n,
     \end{equation}
at the value $s=0$ ($l=-(n+1/2)$): $N^{(0,n,l)}=N_{d}^{(0,n,l)}=1$,
$N^{(1,n-1,l)}=0$,
     \begin{equation}\label{q104}
|d^{(0,n,-(n+1/2))}\rangle = |1/2,-1/2\rangle |n,-n\rangle,
     \end{equation}
while at the values $s=1,...,n$:
$N^{(0,n,l)}=N^{(1,n-1,l)}=N_{c}^{(nl)}=1$,
     \begin{equation}\label{q105}
|v^{(0,n,\l)}\rangle = |1/2,-1/2\rangle |n,l+1/2\rangle,
     \end{equation}
     \begin{equation}\label{q106}
|v^{(1,n-1,l)}\rangle = i|1/2,1/2\rangle |n-1,l-1/2\rangle,
     \end{equation}
     \begin{equation}\label{q107}
\xi^{(nl)} = \sqrt{(n+l+1/2)/6}.
     \end{equation}

For the subspaces of the set $p=1$ (\ref{q47}) with
     \begin{equation}\label{q108}
l = - (n-1/2) + 2s,~s = 0,1,...,n,
     \end{equation}
at the value $s=n$ ($l=(n+1/2)$): $N^{(0,n,l)}=N_{d}^{(0,n,l)}=1$,
$N^{(1,n-1,l)}=0$,
     \begin{equation}\label{q109}
|d^{(0,n,(n+1/2))}\rangle = |1/2,1/2\rangle |n,n\rangle,
     \end{equation}
while at the values $s=0,...,n-1$:
$N^{(0,n,l)}=N^{(1,n-1,l)}=N_{c}^{(nl)}=1$,
     \begin{equation}\label{q110}
|v^{(0,n,\l)}\rangle = |1/2,1/2\rangle |n,l-1/2\rangle,
     \end{equation}
     \begin{equation}\label{q111}
|v^{(1,n-1,l)}\rangle = -i|1/2,-1/2\rangle |n-1,l+1/2\rangle,
     \end{equation}
     \begin{equation}\label{q112}
\xi^{(nl)} = \sqrt{(n-l+1/2)/6}.
     \end{equation}

\subsection{Transitions $J_{0}=1/2\rightarrow J_{1}=3/2$}

For the subspaces of the set $p=0$ (\ref{q47}) with
     \begin{equation}\label{q113}
l = - (n+1/2) + 2s,~s = 0,1,...,n,
     \end{equation}
     \begin{equation}\label{q114}
\xi^{(nl)} = \sqrt{(2n-l-1/2)/12},
     \end{equation}
for all $l$ (\ref{q113}), at the value $s=0$ ($l=-(n+1/2)$):
$N^{(0,n,l)}=N^{(1,n-1,l)}=N_{c}^{(nl)}=1$,
     \begin{equation}\label{q115}
|v^{(0,n,l))}\rangle = |1/2,-1/2\rangle |n,-n\rangle,
     \end{equation}
     \begin{equation}\label{q116}
|v^{(1,n-1,l)}\rangle = i|3/2,-3/2\rangle |n-1,-n+1 \rangle,
     \end{equation}
at the value $s=n$ ($l=n-1/2$):
$N^{(0,n,l)}=N^{(1,n-1,l)}=N_{c}^{(nl)}=1$,
     \begin{equation}\label{q117}
|v^{(0,n,l))}\rangle = |1/2,-1/2\rangle |n,n\rangle,
     \end{equation}
     \begin{equation}\label{q118}
|v^{(1,n-1,l)}\rangle = i|3/2,1/2\rangle |n-1,n-1 \rangle,
     \end{equation}
while at the values $s=1,...,n-1$: $N^{(0,n,l)}=N_{c}^{(nl)}=1$,
$N^{(1,n-1,l)}=2$, $N_{d}^{(1,n-1,l)}=1$,
     \begin{equation}\label{q119}
|v^{(0,n,l)}\rangle = |1/2,-1/2\rangle |n,l + 1/2\rangle,
     \end{equation}
     \begin{equation}\label{q120}
|v^{(1,n-1,\l)}\rangle = c_{0} |w_{0}^{(1,n-1,\l)}\rangle + c_{1}
|w_{1}^{(1,n-1,\l)}\rangle ,
     \end{equation}
     \begin{equation}\label{q121}
|d^{(1,n-1,\l)}\rangle = c_{1} |w_{0}^{(1,n-1,\l)}\rangle - c_{0}
|w_{1}^{(1,n-1,\l)}\rangle ,
     \end{equation}
where
     \begin{equation}\label{q122}
c_{0} =  \sqrt{\frac{3(n-l-1/2)}{2(2n-l-1/2)}},~ c_{1} =
\sqrt{\frac{(n+l+1/2)}{2(2n-l-1/2)}},
     \end{equation}
     \begin{equation}\label{q123}
|w_{k}^{(1,n-1,l)}\rangle = i|3/2,-3/2+2k \rangle |n-1,l+3/2-2k
\rangle .
     \end{equation}

For the subspaces of the set $p=1$ (\ref{q47}) with
     \begin{equation}\label{q124}
l = - (n-1/2) + 2s,~s = 0,1,...,n,
     \end{equation}
     \begin{equation}\label{q125}
\xi^{(nl)} = \sqrt{(2n+l-1/2)/12},
     \end{equation}
for all $l$ (\ref{q124}), at the value $s=0$ ($l=-(n-1/2)$):
$N^{(0,n,l)}=N^{(1,n-1,l)}=N_{c}^{(nl)}=1$,
     \begin{equation}\label{q126}
|v^{(0,n,l))}\rangle = |1/2,1/2\rangle |n,-n\rangle,
     \end{equation}
     \begin{equation}\label{q127}
|v^{(1,n-1,l)}\rangle = i|3/2,-1/2\rangle |n-1,-n+1 \rangle,
     \end{equation}
at the value $s=n$ ($l=n+1/2$):
$N^{(0,n,l)}=N^{(1,n-1,l)}=N_{c}^{(nl)}=1$,
     \begin{equation}\label{q128}
|v^{(0,n,l))}\rangle = |1/2,1/2\rangle |n,n\rangle,
     \end{equation}
     \begin{equation}\label{q129}
|v^{(1,n-1,l)}\rangle = i|3/2,3/2\rangle |n-1,n-1 \rangle,
     \end{equation}
while at the values $s=1,...,n-1$: $N^{(0,n,l)}=N_{c}^{(nl)}=1$,
$N^{(1,n-1,l)}=2$, $N_{d}^{(1,n-1,l)}=1$,
     \begin{equation}\label{q130}
|v^{(0,n,l)}\rangle = |1/2,1/2\rangle |n,l - 1/2\rangle,
     \end{equation}
     \begin{equation}\label{q131}
|v^{(1,n-1,\l)}\rangle = c_{0} |w_{0}^{(1,n-1,\l)}\rangle + c_{1}
|w_{1}^{(1,n-1,\l)}\rangle ,
     \end{equation}
     \begin{equation}\label{q132}
|d^{(1,n-1,\l)}\rangle = c_{1} |w_{0}^{(1,n-1,\l)}\rangle - c_{0}
|w_{1}^{(1,n-1,\l)}\rangle ,
     \end{equation}
where
     \begin{equation}\label{q133}
c_{0} =  \sqrt{\frac{(n-l+1/2)}{2(2n+l-1/2)}},~ c_{1} =
\sqrt{\frac{3(n+l-1/2)}{2(2n+l-1/2)}},
     \end{equation}
     \begin{equation}\label{q134}
|w_{k}^{(1,n-1,l)}\rangle = i|3/2,-1/2+2k \rangle |n-1,l+1/2-2k
\rangle .
     \end{equation}

\subsection{Transitions $J_{0}=3/2\rightarrow J_{1}=1/2$}

For the subspaces of the set $p=0$ (\ref{q47}) with
     \begin{equation}\label{q135}
l = - (n+3/2) + 2s,~s = 0,1,...,n+1,
     \end{equation}
at the values $s=0$ ($l=-(n+3/2)$) and $s=n+1$ ($l=(n+1/2)$:
$N^{(0,n,l)}=N_{d}^{(0,n,l)}=1$, $N^{(1,n-1,l)}=0$,
     \begin{equation}\label{q136}
|d^{(0,n,-(n+3/2))}\rangle = |3/2,-3/2\rangle |n,-n\rangle,
     \end{equation}
     \begin{equation}\label{q137}
|d^{(0,n,(n+1/2))}\rangle = |3/2,1/2\rangle |n,n\rangle,
     \end{equation}
while at the values $s=1,...,n$: $N^{(0,n,l)}=2$,
$N_{d}^{(0,n,l)}=1$, $N^{(1,n-1,l)}=N_{c}^{(nl)}=1$,
     \begin{equation}\label{q138}
|v^{(1,n-1,l)}\rangle = i|1/2,-1/2\rangle |n-1,l + 1/2\rangle,
     \end{equation}
     \begin{equation}\label{q139}
|v^{(0,n,l)}\rangle = c_{0} |w_{0}^{(0,n,l)}\rangle + c_{1}
|w_{1}^{(0,n,l)}\rangle ,
     \end{equation}
     \begin{equation}\label{q140}
|d^{(0,n,l)}\rangle = c_{1} |w_{0}^{(0,n,l)}\rangle - c_{0}
|w_{1}^{(0,n,l)}\rangle ,
     \end{equation}
where
     \begin{equation}\label{q141}
c_{0} =  \sqrt{\frac{3(n+l+3/2)}{2(2n+l+5/2)}},~ c_{1} =
\sqrt{\frac{(n-l+1/2)}{2(2n+l+5/2)}},
     \end{equation}
     \begin{equation}\label{q142}
|w_{k}^{(0,n,l)}\rangle = |3/2,-3/2+2k \rangle |n,l+3/2-2k \rangle,
     \end{equation}
     \begin{equation}\label{q143}
\xi^{(nl)} = \sqrt{(2n+l+5/2)/12}.
     \end{equation}

For the subspaces of the set $p=1$ (\ref{q47}) with
     \begin{equation}\label{q144}
l = - (n+1/2) + 2s,~s = 0,1,...,n+1,
     \end{equation}
at the values $s=0$ ($l=-(n+1/2)$) and $s=n+1$ ($l=(n+3/2)$:
$N^{(0,n,l)}=N_{d}^{(0,n,l)}=1$, $N^{(1,n-1,l)}=0$,
     \begin{equation}\label{q145}
|d^{(0,n,-(n+1/2))}\rangle = |3/2,-1/2\rangle |n,-n\rangle,
     \end{equation}
     \begin{equation}\label{q146}
|d^{(0,n,(n+3/2))}\rangle = |3/2,3/2\rangle |n,n\rangle,
     \end{equation}
while at the values $s=1,...,n$: $N^{(0,n,l)}=2$,
$N_{d}^{(0,n,l)}=1$, $N^{(1,n-1,l)}=N_{c}^{(nl)}=1$,
     \begin{equation}\label{q147}
|v^{(1,n-1,l)}\rangle = i|1/2,1/2\rangle |n-1,l - 1/2\rangle,
     \end{equation}
     \begin{equation}\label{q148}
|v^{(0,n,l)}\rangle = c_{0} |w_{0}^{(0,n,l)}\rangle + c_{1}
|w_{1}^{(0,n,l)}\rangle ,
     \end{equation}
     \begin{equation}\label{q149}
|d^{(0,n,l)}\rangle = c_{1} |w_{0}^{(0,n,l)}\rangle - c_{0}
|w_{1}^{(0,n,l)}\rangle ,
     \end{equation}
where
     \begin{equation}\label{q150}
c_{0} =  \sqrt{\frac{(n+l+1/2)}{2(2n-l+5/2)}},~ c_{1} =
\sqrt{\frac{3(n-l+3/2)}{2(2n-l+5/2)}},
     \end{equation}
     \begin{equation}\label{q151}
|w_{k}^{(0,n,l)}\rangle = |3/2,-1/2+2k \rangle |n,l+1/2-2k \rangle,
     \end{equation}
     \begin{equation}\label{q152}
\xi^{(nl)} = \sqrt{(2n-l+5/2)/12}.
     \end{equation}

\section{Single-photon transitions}

The dimensions of the subspaces $V^{(anl)}$ also do not exceed the
value $n+1$ (\ref{q53}). So the eigenvalues and eigenvectors of
operators $\hat{D}_{a}$ in these subspaces may be easily obtained
analytically in the case of single-photon transitions $n=1$.

For the subspaces with
     \begin{equation}\label{q153}
l = - (J_{0}-1),-J_{0},...,J_{0}-1,
     \end{equation}
$N^{(0,1,l)}=2$, $N_{d}^{(0,1,l)}=1$, $N^{(1,0,l)}=N_{c}^{(1,l)}=1$,
     \begin{equation}\label{q154}
|v^{(0,1,l)}\rangle = c_{0} |w_{0}^{(0,1,l)}\rangle + c_{1}
|w_{1}^{(0,1,l)}\rangle ,
     \end{equation}
     \begin{equation}\label{q155}
|d^{(0,1,l)}\rangle = c_{1} |w_{0}^{(0,1,l)}\rangle - c_{0}
|w_{1}^{(0,1,l)}\rangle ,
     \end{equation}
     \begin{equation}\label{q156}
|v^{(1,0,l)}\rangle = i|J_{1},l \rangle |0,0 \rangle ,
     \end{equation}
where
     \begin{equation}\label{q157}
c_{0} = \frac{\xi_{0}^{(l)}}{\xi^{(l)}},~ c_{1} =
\frac{\xi_{1}^{(l)}}{\xi^{(l)}},~ \xi^{(l)} = \xi^{(1l)} =
\sqrt{\xi_{0}^{(l)2}+\xi_{1}^{(l)2}},
     \end{equation}
     \begin{equation}\label{q158}
\xi_{0}^{(l)} = f_{1}(l-1),~ \xi_{1}^{(l)} = f_{2}(l+1),
     \end{equation}
     \begin{equation}\label{q159}
|w_{k}^{(0,1,l)}\rangle = |J_{0},l-1+2k \rangle |1,1-2k \rangle,~
k=0,1,
     \end{equation}
while $f_{1}(m)$ and $f_{2}(m)$ are defined by the equations
(\ref{q21})-(\ref{q22}).

In the case of transitions $J_{0}=J\rightarrow J_{1}=J-1$:
     \begin{equation}\label{q160}
\xi_{0}^{(l)} =
\left[\frac{(J-l)(J-l+1)}{(2J+1)2J(2J-1)}\right]^{1/2},
     \end{equation}
     \begin{equation}\label{q161}
\xi_{1}^{(l)} =
\left[\frac{(J+l)(J+l+1)}{(2J+1)2J(2J-1)}\right]^{1/2},
     \end{equation}
in the case of transitions $J_{0}=J\rightarrow J_{1}=J$:
     \begin{equation}\label{q162}
\xi_{0}^{(l)} =
\left[\frac{(J+l)(J-l+1)}{(J+1)2J(2J+1)}\right]^{1/2},
     \end{equation}
     \begin{equation}\label{q163}
\xi_{1}^{(l)} = -
\left[\frac{(J-l)(J+l+1)}{(J+1)2J(2J+1)}\right]^{1/2},
     \end{equation}
in the case of transitions $J_{0}=J\rightarrow J_{1}=J+1$:
     \begin{equation}\label{q164}
\xi_{0}^{(l)} =
\left[\frac{(J+l)(J+l+1)}{(2J+3)(2J+2)(2J+1)}\right]^{1/2},
     \end{equation}
     \begin{equation}\label{q165}
\xi_{1}^{(l)} =
\left[\frac{(J-l)(J-l+1)}{(2J+3)(2J+2)(2J+1)}\right]^{1/2}.
     \end{equation}

For the subspaces with $l = \pm J_{0}$ in the case of transitions
$J_{0}=J\rightarrow J_{1}=J-1$: $N^{(1,0,l)}=0$,
$N^{(0,1,l)}=N_{d}^{(0,1,l)}=1$,
     \begin{equation}\label{q166}
|d^{(0,1,\pm J)}\rangle = |J_{0},\pm (J-1) \rangle |1,\pm 1\rangle ,
     \end{equation}
in the case of transitions $J_{0}=J\rightarrow J_{1}=J$:
$N^{(1,0,l)}=N^{(0,1,l)}=N_{c}^{(1,l)}=1$,
     \begin{equation}\label{q167}
|v^{(0,1,\pm J)}\rangle = \pm |J_{0},\pm (J-1) \rangle |1,\pm
1\rangle ,
     \end{equation}
     \begin{equation}\label{q168}
|v^{(1,0,\pm J)}\rangle =  i |J_{1},\pm J \rangle |0,0\rangle ,
     \end{equation}
     \begin{equation}\label{q169}
\xi^{(l)} = \frac{1}{\sqrt{(J+1)(2J+1)}},
     \end{equation}
in the case of transitions $J_{0}=J\rightarrow J_{1}=J+1$:
$N^{(1,0,l)}=N^{(0,1,l)}=N_{c}^{(1,l)}=1$,
     \begin{equation}\label{q170}
|v^{(0,1,\pm J)}\rangle = |J_{0},\pm (J-1) \rangle |1,\pm 1\rangle ,
     \end{equation}
     \begin{equation}\label{q171}
|v^{(1,0,\pm J)}\rangle =  i |J_{1},\pm J \rangle |0,0\rangle ,
     \end{equation}
     \begin{equation}\label{q172}
\xi^{(l)} = \sqrt{\frac{J}{(J+1)(2J+3)}}.
     \end{equation}

For the subspaces with $l = \pm (J_{0}+1)$ in the case of
transitions $J_{0}=J\rightarrow J_{1}=J-1$ and $J_{0}=J\rightarrow
J_{1}=J$: $N^{(1,0,l)}=0$, $N^{(0,1,l)}=N_{d}^{(0,1,l)}=1$,
     \begin{equation}\label{q173}
|d^{(0,1,\pm (J+1))}\rangle = |J_{0},\pm J \rangle |1,\pm 1\rangle ,
     \end{equation}
in the case of transitions $J_{0}=J\rightarrow J_{1}=J+1$:
$N^{(1,0,l)}=N^{(0,1,l)}=N_{c}^{(1,l)}=1$,
     \begin{equation}\label{q174}
|v^{(0,1,\pm (J+1))}\rangle = |J_{0},\pm J \rangle |1,\pm 1\rangle ,
     \end{equation}
     \begin{equation}\label{q175}
|v^{(1,0,\pm (J+1))}\rangle =  i |J_{1},\pm (J+1) \rangle
|0,0\rangle ,
     \end{equation}
     \begin{equation}\label{q176}
\xi^{(l)} = \frac{1}{\sqrt{2J+3}}.
     \end{equation}

Let us consider an example, when initially there is a single photon
with the $\sigma^{+}$ polarization $\textbf{s}_{1}$ in the cavity,
while the atom is at its ground level, so that the initial density
matrix of the system is as follows:
     \begin{equation}\label{q177}
\hat{\rho}(0) = \sum_{m=-J_{0}}^{J_{0}} n^{(0)}_{m} |J_{0},m \rangle
|1,1 \rangle \langle 1,1| \langle J_{0},m|.
     \end{equation}
The probability to find in the cavity at the instant of time $t$ the
photon with the orthogonal $\sigma^{-}$ polarization
$\textbf{s}_{2}$ is given by the formula:
     \begin{equation}\label{q178}
w(t) = tr\{\hat{\rho}(t)|1,-1 \rangle \langle 1,-1|\}.
     \end{equation}
Then from the equations (\ref{q55})-(\ref{q65}) with an account of
(\ref{q153})-(\ref{q176}) we obtain:
     \begin{equation}\label{q179}
w(t) = \sum_{m=-J_{0}}^{J_{0}} n^{(0)}_{m} |F_{m+1}(t)|^{2},
     \end{equation}
where
     \begin{equation}\label{q180}
F_{m}(t) = \frac{\xi^{(m)}_{0}\xi^{(m)}_{1}}{\xi^{(m)2}} \left[
C^{(m)}(t) - e^{i\Delta t /2} \right],
     \end{equation}
     \begin{equation}\label{q181}
C^{(m)}(t) = \cos\left(\frac{\Omega^{(m)} t}{2}\right) + i
\frac{\Delta}{\Omega^{(m)}} \sin\left(\frac{\Omega^{(m)}
t}{2}\right),
     \end{equation}
     \begin{equation}\label{q182}
\Omega^{(m)} = \sqrt{\Delta^{2} + \theta^{2}\xi^{(m)2}},
     \end{equation}
while $\xi^{(m)}_{0}$, $\xi^{(m)}_{1}$ and $\xi^{(m)}$ are defined
by (\ref{q157})-(\ref{q158}). In the case of the exact resonance
$\Delta =0$ and the atom initially prepared at the pure state
$n^{(0)}_{m}=\delta_{m,-1}$:
     \begin{equation}\label{q183}
w(t) = \frac{1}{4}\left[1-\cos\left(\frac{\Omega^{(0)}t}{2}
\right)\right]^{2}, \Omega^{(0)}=\theta f_{2}(1)\sqrt{2},
     \end{equation}
so that this probability attains the unity value at the instant of
time $t=2\pi/\Omega^{(0)}$.

\section{Single-photon emission}

Now let us consider the emission of the photon by an atom at excited
state into an empty cavity, so that the initial density matrix of
the system is as follows:
     \begin{equation}\label{q184}
\hat{\rho}(0) = \sum_{m,m'=-J_{1}}^{J_{1}} n^{(1)}_{mm'} |J_{1},m
\rangle |0,0 \rangle \langle 0,0| \langle J_{1},m'|.
     \end{equation}
While considering the photon emission let us take into account the
relaxation processes. Then in the equation (\ref{q54}) for the
system density matrix the relaxation terms must be added:
     \begin{equation}\label{q185}
\frac{d\hat{\rho}}{dt} = \frac{i}{2}
\left[\hat{\Omega},\hat{\rho}\right] + \hat{L}_{c} + \hat{L}_{a},
     \end{equation}
where the term
     \begin{equation}\label{q186}
\hat{L}_{c} = -\frac{\gamma_{c}}{2}\sum_{i=1}^{2} \left(
\hat{a}_{i}^{\dag}\hat{a}_{i}\hat{\rho} +
\hat{\rho}\hat{a}_{i}^{\dag}\hat{a}_{i} -
2\hat{a}_{i}\hat{\rho}\hat{a}_{i}^{\dag} \right),
     \end{equation}
describes the relaxation of the cavity field with the rate
$\gamma_{c}$, while the term
     \begin{equation}\label{q187}
\hat{L}_{a} = -\frac{\gamma_{a}}{2}(2J_{1}+1)\sum_{i=1}^{3} \left(
\hat{g}_{i}^{\dag}\hat{g}_{i}\hat{\rho} +
\hat{\rho}\hat{g}_{i}^{\dag}\hat{g}_{i} -
2\hat{g}_{i}\hat{\rho}\hat{g}_{i}^{\dag} \right),
     \end{equation}
describes the spontaneous emission of the atom at the transitions
$J_{1}\rightarrow J_{0}$ into a free-space modes with the rate
$\gamma_{a}$. In the equation (\ref{q186}) the summation is carried
out over two cavity modes with unit polarization vectors
$\textbf{s}_{1}$ ($\sigma^{+}$) and $\textbf{s}_{2}$ ($\sigma^{-}$),
while in the equation (\ref{q187}) the summation is carried out over
all three unit polarization vectors $\textbf{s}_{1}$ ($\sigma^{+}$),
$\textbf{s}_{2}$ ($\sigma^{-}$) and $\textbf{s}_{3}$ ($\pi$),
     \begin{equation}\label{q188}
\sum_{i=1}^{2} \hat{a}_{i}^{\dag}\hat{a}_{i} = \hat{n},~~~
\sum_{i=1}^{3} \hat{g}_{i}^{\dag}\hat{g}_{i} =
\hat{P}_{A}^{(1)}/(2J_{1}+1),
     \end{equation}
$\hat{n}$ and $\hat{P}_{A}^{(1)}$ being the photon number operator
and the projector on the subspace of states referring to the excited
atomic level. The presence of thermal photons in the relaxation
terms is neglected since the photon emission in the optical domain
is considered.

The probability $w$ to find the photon in the cavity and photon
polarization matrix $\hat{\sigma}$ are described by the $2\times 2$
hermitian density matrix of the photon $\hat{w}$:
     \begin{equation}\label{q189}
w = \sum_{q=\pm 1} w_{qq},~~~ \sigma_{qq'} = w_{qq'}/w,
     \end{equation}
     \begin{equation}\label{q190}
w_{qq'} = tr\left\{\hat{\rho} |1,q' \rangle \langle 1,q| \right\},
     \end{equation}
where $\hat{\rho}$ is the system density matrix.

The orthonormal basis of the space of system states, which
contribute to the photon dynamics, consists of the eigenvectors of
the interaction operator:
     \begin{equation}\label{q191}
|u^{(m)}_{1}\rangle  = c^{(+)}_{m} |v^{(0,1,m)}\rangle + c^{(-)}_{m}
|v^{(1,0,m)}\rangle,
     \end{equation}
     \begin{equation}\label{q192}
|u^{(m)}_{2}\rangle  = c^{(-)}_{m} |v^{(0,1,m)}\rangle - c^{(+)}_{m}
|v^{(1,0,m)}\rangle,
     \end{equation}
     \begin{equation}\label{q193}
c^{(\pm)}_{m} =\sqrt{\frac{1}{2}\left(1 \pm
\frac{\Delta}{\Omega^{(m)}}\right)},
     \end{equation}     \
where $|v^{(0,1,m)}\rangle$, $|v^{(1,0,m)}\rangle$, and
$\Omega^{(m)}$ are defined by (\ref{q154}), (\ref{q156}) and
(\ref{q182}) correspondingly. The master equation
(\ref{q185})-(\ref{q187}) for the system density matrix elements $
\rho^{mm'}_{kk'} = \langle u_{k}^{(m)} | \hat{\rho} |u_{k'}^{(m')}
\rangle$, ($k,k'=1,2$) in the basis of these states is as follows:
     \begin{equation}\label{q194}
\frac{d}{dt} \rho^{mm'}_{kk'} = \frac{i}{2} \left( \lambda_{k}^{m} -
\lambda_{k'}^{m'} \right) \rho^{mm'}_{kk'} + L^{mm'}_{kk'},
     \end{equation}
     \begin{equation}\label{q195}
L^{mm'}_{kk'} = - \frac{1}{2} \sum_{s=1}^{2} \left( \gamma^{m}_{ks}
\rho^{mm'}_{sk'} + \rho^{mm'}_{ks} \gamma^{m'}_{sk'}\right),
     \end{equation}
where
     \begin{equation}\label{q196}
\lambda_{1}^{m} = \Omega^{(m)},~~~ \lambda_{2}^{m} = -\Omega^{(m)},
     \end{equation}
     \begin{equation}\label{q197}
\gamma^{m}_{11} = \gamma_{c} c_{m}^{(+)2} + \gamma_{a} c_{m}^{(-)2},
\gamma^{m}_{22} = \gamma_{a} c_{m}^{(+)2} + \gamma_{c} c_{m}^{(-)2},
     \end{equation}
     \begin{equation}\label{q198}
\gamma^{m}_{12} = \gamma^{m}_{21} =
c_{m}^{(+)}c_{m}^{(-)}(\gamma_{c} - \gamma_{a}).
     \end{equation}
In the case of strong coupling ($\Omega^{(m)} \gg
\gamma_{c},\gamma_{a}$) the solution of the equations
(\ref{q194})-(\ref{q195}) is easily obtained:
     \begin{equation}\label{q199}
\rho^{mm'}_{kk'}(t) = \rho^{mm'}_{kk'}(0) f_{mk}^{*}(t) f_{m'k'}(t),
     \end{equation}
where
     \begin{equation}\label{q200}
f_{mk}(t) = \exp \left\{ -\frac{1}{2} \left( \gamma_{kk}^{m} + i
\lambda_{k}^{m} \right) t \right\}.
     \end{equation}
Then we obtain the following expression for the photon density
matrix (\ref{q190}) in this approximation:
     \begin{equation}\label{q201}
w_{q'q} = e^{-\gamma t} \sum_{m,m'} n^{(1)}_{mm'} h_{mq}^{*}
h_{m'q'} \delta_{m'-m,q'-q},
     \end{equation}
     \begin{equation}\label{q202}
h_{qm} = b_{q}^{(m)}\frac{\theta}{\Omega^{(m)}} \left( p_{m} + i
q_{m} \right),
     \end{equation}
     \begin{equation}\label{q203}
p_{m} = \sinh\left(\frac{\Delta^{(m)} t}{2}\right)
\cos\left(\frac{\Omega^{(m)}t}{2}\right),
     \end{equation}
     \begin{equation}\label{q204}
q_{m} = \cosh\left(\frac{\Delta^{(m)} t}{2}\right)
\sin\left(\frac{\Omega^{(m)}t}{2}\right),
     \end{equation}
     \begin{equation}\label{q205}
\gamma = \frac{(\gamma_{c}+\gamma_{a})}{2},~ \Delta^{(m)}
=\frac{\Delta}{\Omega^{(m)}} \frac{(\gamma_{c}-\gamma_{a})}{2},
     \end{equation}
where $b_{1}^{(m)}=\xi_{0}^{(m)}$, $b_{-1}^{(m)}=\xi_{1}^{(m)}$,
while $\xi_{0}^{(m)}$, $\xi_{1}^{(m)}$, and $\Omega^{(m)}$ are
defined by (\ref{q158}) and (\ref{q182}) correspondingly.

For example, in the case of transitions $J_{0}=0\rightarrow J_{1}=1$
we obtain from (\ref{q201})-(\ref{q205}) the following expressions
for the photon polarization matrix $\hat{\sigma}$ and the
probability to find the photon into the cavity $w(t)$ (\ref{q189}):
     \begin{equation}\label{q206}
\sigma_{qq'} = n^{(1)}_{qq'}/n_{0},~ n_{0} = n^{(1)}_{11} +
n^{(1)}_{-1,-1},
     \end{equation}
     \begin{equation}\label{q207}
w(t) = n_{0} \frac{\theta^{2}\xi^{(1)2}}{\Omega^{(1)2}} F(t),~
\xi^{(1)} = \frac{1}{\sqrt{3}},
     \end{equation}
     \begin{equation}\label{q208}
F(t) = \sin^{2}\left(\frac{\Omega^{(1)}t}{2}\right) +
 \sinh^{2}\left(\frac{\Delta^{(1)} t}{2}\right),
     \end{equation}
where $\Omega^{(1)}$ and $\Delta^{(1)}$ are defined by (\ref{q182})
and (\ref{q205}).

\section{Conclusions}

The problem of diagonalization of the hamiltonian of the
Jaynes-Cummings model with degenerate atomic levels and
polarization-degenerate field mode may be reduced to the
diagonalization of hermitian matrix which dimension does not exceed
the degree of degeneracy of atomic levels. Such diagonalization may
be easily performed analytically for low values ($J_{0},J_{1} \leq
3/2$) of atomic angular momentum or numerically otherwise. The
evolution operator of the system is decomposed into blocks
determined by the photon number $n$ and the projection $l$ of the
system angular momentum on the cavity axis, such blocks being
expressed through the corresponding eigenvalues and eigenvectors of
the system hamiltonian. The obtained expressions may be used for
analytical or numerical studies of the atom-field dynamics in
micro-cavities with arbitrary angular momenta of atomic levels and
arbitrary polarization of the field mode.

\section*{Data availability}

Data sharing is not applicable to this article as no new data were
created or analyzed in this study.

\section*{Author Declarations}
\subsection*{Conflict of interest}

The author have no conflicts to disclose.

\end{document}